\documentclass[conference]{IEEEtran}

\usepackage{cite}
\usepackage{graphicx}
\usepackage{amsmath}
\usepackage[hidelinks]{hyperref}
\usepackage{xcolor}
\usepackage{enumitem}

\title{Road Rules for Radio: Why Your Wi-Fi Got Better \\ [1ex] \large An accessible review of the mechanisms behind faster, fairer, more reliable wireless networks}

\author{\IEEEauthorblockN{Bradley Fang, Michael Roger}
\IEEEauthorblockA{bradleyfang@gmail.com, michael.roger@columbia.edu}
}

\begin{document}
\maketitle

\begin{abstract}
WiFi allows for the connection of devices and people around the globe. It has proven to be a monumental and revolutionary tool that keeps the world connected. However, recent WiFi advancements are numerous and at times confusing. WiFi has grown significantly over the years, yet few understand the scope and scale of WiFi progression as a whole. This paper tackles that problem, providing a broad literature review on the advancements of key WiFi features to date. This paper will center on seven key areas of focus: (1) bandwidth, (2) battery life, (3) traffic collisions, (4) interference, (5) data-intensive transmissions, (6) numerous devices, and (7) peak throughput/modulation. Each section will focus on WiFi’s problems, how those problems were fixed, as well as the limitations of existing solutions. Moreover, the paper explains the role of new unreleased technologies in these seven areas. This includes exploring the upcoming WiFi 8 standard based on the IEEE 802.11bn “Ultra High Reliability” (UHR) specification and how it builds upon current specifications. Compared to previous specifications, WiFi 8 marks a stronger and more significant shift toward prioritizing reliability over pure data rates. Beyond a sole literature review, this paper uses a novel analogy. A road/highway analogy will be integrated throughout the paper to facilitate understanding of networking mechanisms. This paper is approachable and is written such that someone with very little WiFi knowledge should come away with a strong understanding of WiFi. As is typical of literature review papers, technical claims will be grounded in prior work.
\end{abstract}

\section*{1 INTRODUCTION}
WiFi in the original IEEE 802.11 standard solely used frequencies in the 2.4GHz frequency band, and thus had extremely limited bandwidth and throughput potential. This specification had a max theoretical speed of 11Mbps. Today’s commercially available WiFi 7 (802.11be) has a maximum theoretical speed of 46Gbps; that is an increase of over 4000x. 
One may wonder how WiFi has arrived at this current state after numerous developments and specifications of WiFi. However, most attempts at understanding WiFi broadly are convoluted and intensely time-consuming. Much of the existing research does not highlight the true benefits or downsides of WiFi, or covers WiFi advancements in such a way that very few are able to understand it. The objective of this paper is to unify these concepts in a balance of readability and conceptual understanding such that both a layman and an expert in WiFi may come away having a stronger knowledge of the developments of WiFi to date. In order to facilitate comprehension from a broader audience, specific WiFi jargon will not be assumed as common knowledge and all major concepts will be distilled to a readable and comprehensible form.
Although the developments may seem disparate at a glance, they can become easily understood when connected through a common link. In this paper, the common link is an analogy, connecting major WiFi concepts to the various functions of a highway. Each concept will be explored thoroughly, including the problems that WiFi had to solve and how those problems were addressed with WiFi features. Then, the limitations and trade-offs of each approach will be addressed. The concepts are addressed in the following order: 
First, bandwidth availability and its effect on throughput. 
Second, battery life and the progression of battery-saving mechanisms. 
Third, collision avoidance frameworks to avoid network retransmissions.
Fourth, interference in wide network channels and mitigating its impact.
Fifth, data-intensive applications and a limited number of links.
Sixth, routing network traffic to multiple users simultaneously.
Seventh, achieving peak throughput through modulation. Additionally, many of these concept explorations will include new specifications from the upcoming WiFi 8 (IEEE802.11be specification). With WiFi 8, there has been a concerted effort toward reliability over raw throughput gains (e.g., reducing the latency for 95th latency distribution by 25\%). \cite{IEEE80211bn} As such, the new specifications of WiFi 8 reflect this focus on reliability.

\section*{2.1 Bandwidth}
Bandwidth is the usable frequency spectrum available for sending data. \cite{BE22} Bandwidth is one of the key elements of a connection that bounds the ceiling for potential throughput. \cite{CB11} With the WiFi 1 (802.11b) specification that is widely available to consumers, \cite{WLAN01} only the 2.4GHz band is used. \cite{WLAN01,BT03} As such, there is only a small subsection (typically 2400MHz-2483.5MHz in US) that is available for data transmission. \cite{Shabani13} Viewing this problem through the road analogy, this is like having limited land space allocated for building roads for cars. The key question is: with limited land, how can the potential number of cars on the road be maximized?

\subsection*{2.1.1 Total Bandwidth}
The 2.4GHz band has a limited frequency range, limiting total throughput potential across all users. \cite{WLAN01} As such, an obvious solution would be to simply use a wider range of frequencies for transmissions.
In the United States, the FCC has allowed a wider range of frequencies to be used for WiFi. \cite{FCC20} In Wi-Fi 2-era deployments, 5 GHz operation (5150-5850 MHz, excluding 5350-5470 MHz) was added to the IEEE standard (802.11a and later). \cite{WiFi6Base} WiFi 6E followed this development, adding the 6GHz band (5925-7125MHz) after it was opened for unlicensed use in the United States in 2020 \cite{FCC20,6E25}. These three frequency bands are like three separate highway systems: slow backroads (2.4GHz), medium capacity interstate highways (5GHz), and high capacity super-highways (6GHz). When combining the total bandwidth from all three, the total allocated spectrum for these bands in the U.S. went from just 83.5MHz in early specifications to now 83.5 + 580 + 1200 = 1863.5MHz. This greatly increases aggregate throughput potential \cite{FCCFactSheet20,6E25}.

\subsection*{2.1.2 Wider Channels}
Just like combining two adjacent highways together into one, Wi-Fi has been able to support wider and wider channels via channel bonding \cite{CB11}. Very wide channel widths (up to 320 MHz) have been implemented in the recent WiFi implementation IEEE 802.11be (Wi-Fi 7) \cite{BE22,Deng20}. Since each client can only connect to one channel at a time \cite{Ergen02} (before MLO, Section 2.5), the width of the channel determines the upper ceiling for that client’s transmission potential. \cite{CB11} For an individual WiFi user, a 320 MHz channel offers 16 times the bandwidth potential as a 20 MHz channel. \cite{BE22} With the analogy, this individual user can be considered an online shopper who has ordered items from Amazon.  The box of items is analogous to a network packet that contains data.  By using a wider highway, the online shopper has the potential to receive many more boxes at a time.

\subsection*{2.1.3 Limitations}
Through channel bonding, wider channels permit significantly higher throughput rates. \cite{CB11} However, wider channels are not necessarily beneficial in all contexts. \cite{CB11} High throughput gained through wider channels is significantly more vulnerable to interference simply because the channel takes up a wider chunk of the frequency spectrum. \cite{Autoconfig} This can be thought of as an extra-wide highway having higher susceptibility to accidents, and each accident has greater residual effects compared to an accident on a smaller highway.
When compared under similar indoor conditions, higher frequencies (such as the 5GHz and 6GHz bands) often experience high penetration loss through obstructions (walls, floor, etc) \cite{ITU1238}. As a result, 5/6 GHz bands can often provide higher peak throughput, \cite{WiFi6Base,BE22} while 2.4 GHz often remains useful for long-distance coverage. \cite{PL17,CiscoRFDesign} This can be thought of as ultra-wide superhighways that aren’t very long and can’t route well around buildings (6GHz) compared to much thinner backroads that go extremely far and can wind around trees and towns (2.4GHz).
Another limitation to consider is the component of crowding. \cite{MW24} The 2.4 GHz band is often overpopulated by existing devices and faces interference from other 2.4 GHz devices (e.g., Bluetooth and microwave ovens) and neighboring wireless networks \cite{McKayMasuda03VoIP,MW24}. As such, 2.4GHz channels oftentimes have strong coverage potential but high interference. \cite{BT03,MW24} The 5GHz channels also face interference other wireless networks, \cite{WiFi6Base} but these two are in contrast to the 6GHz band which is relatively new; \cite{BE22,6E25} as widespread adoption increases, so will congestion in this frequency band. \cite{BE22}

\section*{2.2 Battery Life}
When data is being transmitted, a device must be awake in order to successfully receive packets, which drains said device’s battery in the process \cite{PSM05}. If no system is in place to protect the battery life of a device, they will lose battery charge at an alarming rate. \cite{PSM05} Through the road analogy, this can be thought of as an online shopper waiting for a time-sensitive package but being unaware of when it will arrive; as a result, the customer simply waits at the front door for the mail man, draining their “battery” (losing energy and sacrificing restful sleep time).

\subsection*{2.2.1 PSM}
Legacy PSM (Power-Saving Mode) was the initial solution to this problem, defined in legacy WiFI (802.11) \cite{PSM05}. The premise is that a device sleeps when it is not transmitting or receiving data. \cite{PSM05} The device will periodically wake to check whether the wireless Access Point (AP) has packets buffered for it to receive \cite{PSM05}. To share this packet buffer status, the AP will send out regularly scheduled notifications spaced around time intervals known as beacon intervals; the common default beacon interval is 100 TUs (102.4 ms), though it is configurable \cite{100TU}. The AP stores packets temporarily and transmits buffered traffic when the device wakes and indicates readiness \cite{PSM05}.
This is like an online shopper checking a remote status board that updates every set interval (e.g., 100ms). If the customer sees on the board that a package is to be delivered, then the customer arranges the delivery with a delivery company. If there are no packages for the customer, then the customer goes back to sleep and will wake again in 100ms in order to check the status board for new updates. This solution is a significant improvement over a system with no energy management at all, but it’s far from the best solution. The system provides enough time for the customer to get microsleeps between each interval before checking the board, but restful sleep for a longer duration could provide better battery life benefits.

\subsection*{2.2.2 TWT}
Beyond PSM, another battery-life saving mechanism is Target Wake time or TWT. \cite{TWT20} TWT originated from WiFi HaLow (IEEE 802.11ah) \cite{TWT20} which focuses on high battery-life Internet of Things (IoT) devices (e.g. smart refrigerators, smart lightbulbs, etc). It was then adapted for Wi-Fi 6 (IEEE 802.11ax) \cite{Khorov19}. With TWT, the device negotiates a wake schedule with the AP \cite{TWT20}. When the agreed upon time comes, the device wakes and transmits/receives data only when needed, saving energy \cite{TWT20}. With TWT, devices can sleep for long periods (e.g., hours or longer) depending on their transmission requirements and negotiated sleep intervals. \cite{TWT20} Using the analogy, this is equivalent to establishing a delivery window for each Amazon customer. Customers do not need to check the delivery status board. Instead, customers can negotiate delivery times with the delivery company, requiring customers to only be awake during the agreed-upon delivery time. Legacy PSM has rest times that are often tied to the AP beacon intervals, \cite{Tabrizi12} while TWT has varying rest times that can be significantly longer depending on application. \cite{TWT20} Not only are these rests longer, but they are also more targeted, allowing the device to wake up only when necessary. \cite{TWT20} In addition, the AP packet buffering is finite; if excess packets are held too long, buffers can overflow and packets may be dropped. \cite{PSM05} TWT’s negotiated scheduling can be calibrated to match real traffic needs and limit overflow while PSM is only based around beacon intervals. \cite{TWTvsPSM}

\subsection*{2.2.3 Limitations \& Future Work}
Wi-Fi PSM typically involves shorter intervals (e.g., hundreds of milliseconds) for standard device traffic. \cite{PSM05} However, cellular PSM designed specifically for IoT devices allows for much longer "deep sleep" periods (e.g., 310 hours in Narrowband IoT). \cite{NBIoT25,NBIoT18,NBIoT17}
An important caveat to note with TWT is that increasing rest time reduces energy expenditures only if the service period does not increase proportionally; \cite{TWT20} otherwise the client simply wakes less often but stays awake longer each time. The average energy use is proportional to the fraction of time the device is awake.

TWT was initially designed for low-power IoT devices; \cite{TWT20} this design purpose contributes to why TWT suffers in high density environments with diverse traffic types. \cite{TWT20} An AP with TWT is incredibly efficient in managing its own devices, but there is no synchronization across other APs which leads to frequent interference. \cite{TWT20}
With WiFi 8, specifications point toward battery saving mechanisms that will save power not with the devices, but rather with the AP. \cite{APPS25} Some AP-specific power-saving features have already been outlined and are likely to be featured in WiFi 8. \cite{APPS25}

\section*{2.3 Traffic Collisions}
Without coordination, Wi-Fi can suffer collisions and re-transmissions because multiple devices attempt to transmit data on the same link at the same time \cite{Bianchi00,Kleinrock75}. How can the link be shared effectively?
There are two intuitive methods.

\subsection*{2.3.1 Time sharing (TDMA-style):}
One method is time sharing (see limitations). Users can alternate transmissions in predetermined time slots in a system called Time Division Multiple Access (TDMA). \cite{Rappaport96} The first user is allotted a fixed time interval to transmit, then the next user is allotted the next interval. \cite{Rappaport96} In terms of highways, each customer is designated a specific time slot to use the road; within that time slot, the customers reserve the entire highway for exclusive use. Customer A gets 12:00am–6:00am; Customer B gets 6:00am–12:00pm, and so on. This style of bandwidth allotment can provide predictable access in some systems, and is oftentimes used in systems where consistent, uninterrupted transmissions are key. \cite{Rappaport96} One such example is in air-traffic control systems where voice intercom quality, and thus a reliable connection, is crucial \cite{Studenberg04}. However, a device that doesn’t have data to transmit when it has a time slot reserved is effectively underutilizing the link. \cite{Rappaport96}

\subsection*{2.3.2 Carrier sensing (CSMA).}
In Carrier Sense Multiple Access (CSMA), a device checks whether the channel is idle by detecting radio signals that indicate whether data is currently being transferred. \cite{Kleinrock75} If busy, the device waits and tries later; if “idle”, it attempts to transmit. \cite{Kleinrock75} This is like a car waiting to merge onto a highway. When the driver of the car sees that the road is safe, they will attempt to merge onto the highway and deliver packages to their customers. Multiple devices may detect the channel as idle at the same time, and if a collision occurs then the device will attempt to listen to the channel again after a random interval. \cite{Bianchi00} The interval width is initially between 0 and CWmin (for example 31 in 802.11b and 15 in 802.11a/g) and doubles after every collision, up to a CWmax, commonly 1023 (or 2\textasciicircum{}10 - 1). \cite{Ergen02,Bianchi00} Each slot time is 9µs (802.11a/WiFi 2 and beyond) \cite{Ergen02}. An important observation is that like TDMA, CSMA only permits one user to transmit at a time. \cite{Bianchi00} This means that after merging onto a highway, one customer’s car will take up the entire highway. \cite{Bianchi00}

\subsection*{2.3.3 OFDMA}
The key limitation of both CSMA and TDMA is that only one user can transmit at one time. \cite{Bianchi00} This greatly limits the stability of devices’s WiFi connections and causes bandwidth underutilization. \cite{Bellalta16} Orthogonal Frequency Division Multiple Access (OFDMA) solves this problem. \cite{Khorov19} OFDMA combines the time-division concept of TDMA and the idle-channel checking of CSMA into a novel idea: transmit data for multiple users simultaneously. OFDMA was introduced in Wi-Fi 6 (802.11ax) \cite{Khorov19} and divides the channel by frequency into smaller chunks called Resource Units (RUs). \cite{Khorov19} Each RU can deliver data to a different client, \cite{Khorov19} allowing transmission to multiple devices at the same time. \cite{Khorov19} This is like splitting up the highway into lanes that cars can use both individually and simultaneously to travel on. OFDMA is split by time \cite{Khorov19} because all Resource Units within a time frame are synchronized and transmitted at the same time. \cite{Khorov19} OFDMA also borrows the CSMA backoff; \cite{Khorov19} OFDMA adapts backoff times to channel conditions and RU availability, \cite{Kim21OFDMARU} but it also includes the random backoff mechanism of legacy CSMA. \cite{Bianchi00}
The greatest benefit of OFDMA is that bandwidth can be used more efficiently; \cite{Khorov19} many clients do not need the entirety of a link at all times \cite{Daldoul20IoT}. This is effective for periodic, predictable, and low-data transmissions such as Internet of Things (IoT) traffic \cite{Daldoul20IoT} as well as for high data transmissions, \cite{Khorov19} since data can be sent on multiple RUs at the same time \cite{Khorov19}. In practical terms, this helps both “always on” traffic (e.g., a video call) transmit alongside bursty downloads cleanly, allowing both to achieve their designated goals. \cite{XuTDMA}.

\subsection*{2.3.4 Limitations \& Future Work}
One important caveat: WiFi does not actually use TDMA. \cite{Ergen02} Because WiFi tends to have bursty and unpredictable transmissions, it becomes difficult to synchronize with TDMA’s strict time slots. \cite{XuTDMA} WiFi has always used a form of CSMA, \cite{Bianchi00} since CSMA’s versatile collision avoidance and random backoff is more suited to the bursty nature of WiFi. \cite{Bianchi00} Regardless, scheduled-access ideas like TDMA are present in and provide context to later developments such as OFDMA \cite{Khorov19}.
With WiFi 6, a client can only be assigned one RU at a time. \cite{Khorov19} This could leave sections of the spectrum unused. With WiFi 7, this problem is addressed; clients are allowed multiple RUs, which can fill unused spectrum space instead of only using singular continuous RU blocks. \cite{Deng20} For WiFi 8, the focus shifts toward interference from other APs rather than more optimization within the same connection. \cite{He25}  As such, WiFi 8 is exploring multi-AP coordination, where APs coordinate signals to avoid interference and sustain strong, consistent signals. \cite{He25}

\section*{2.4 Interference}
WiFi has always needed to contend with interference from other signals. \cite{Gollakota11} Bluetooth devices and microwave ovens are well-known interference sources within the 2.4 GHz frequency space \cite{Golmie03,Nassar11}, which greatly affect the 2.4GHz WiFi frequency band. \cite{Gollakota11} In the 5GHz band, interference often comes from other wireless networks. \cite{Bellalta16} In the newly acquired 6 GHz band, coexistence with incumbents such as fixed-satellite services are sources of interference \cite{Mitsuishi24}. Regardless of which frequency band is used, interference is a pervasive problem in WiFi. \cite{Gollakota11}

\subsection*{2.4.1 Preamble Puncturing}
Preamble puncturing (part of IEEE 802.11ax and extended in IEEE 802.11be \cite{Puncture23,Deng20}) is a feature that allows for the exclusion of subchannels affected by interference while still allowing the rest of the channel to be used \cite{Puncture23}. Each subchannel is the same as a standard channel: 20 MHz \cite{Puncture23,Ergen02}. For example, if a Wi-Fi 7 device aims to use 160 MHz but two 20 MHz subchannels have interference, it can stop transmitting on those two 20 MHz subchannels and still use the remaining 120 MHz \cite{Puncture23}. This is like shutting down an eight lane highway for construction. The solution before preamble puncturing was to shut down the entire highway. With preamble puncturing, only the affected part of the highway is closed.
Preamble puncturing is extremely useful because as channels become wider, they become more susceptible to interference \cite{Puncture23}. Wider channels are directly proportional to peak throughput \cite{Deng20}. Preamble puncturing acts as a safety net for these channels, mitigating much of the risk that inherently comes with wider channels. \cite{Puncture23}

\subsection*{2.4.2 Limitations \& Future Work}
Preamble puncturing is only needed when wider channels are used. \cite{Puncture23} Each puncture closes a 20MHz subchannel. \cite{Puncture23} For future improvements with preamble puncturing, even finer granularity could be sought. Instead of closing entire 20MHz subchannels, a potential solution could be closing only the Resource Units that are affected, allowing usage for thinner channel sizes and saving more bandwidth to be used for transmission. This is currently not possible because puncturing is limited to a minimum size of 20 MHz chunks \cite{Puncture23}, and attempting it would require significantly more complex and detailed interference estimations to match RU granularity. Smaller-scale puncturing is not currently a feature of the WiFi 8 specification. However, for future developments beyond WiFi 8, especially as WiFi shifts toward reliability and loss prevention \cite{He25}, this could be an area for consideration.

\section*{2.5 Large Transmissions}
As the internet devices become more data-intensive, transmission data requirements increase and applications begin demanding higher throughput and lower latency (e.g., high-quality live-streaming, cloud gaming, VR/AR). \cite{Deng20} However, before WiFi 7, devices are able to connect to WiFi APs using only one channel at a time, and the maximum throughput one can hope to achieve is limited by the bandwidth of that channel. \cite{Ergen02,Deng20} In the analogy, this is like allowing a multi-package Amazon delivery to use only one highway system (backroads, highway, super-highway) at a time. How can this problem be addressed?

\subsection*{2.5.1 MLO}
Recently in the IEEE 802.11be (Wi-Fi 7) specification, Multi-Link Operation (MLO) was introduced in order to target the increasingly high throughput demands of newer applications \cite{Deng20}. MLO allows an AP to transmit traffic across multiple WiFi bands (2.4/5/6 GHz) simultaneously \cite{Deng20}. This allows transmission to occur on channels across multiple bands without being constrained to a single channel on one band at a time \cite{Deng20}. In the analogy, this is like sending delivery vans on all available roads, highway systems to deliver packages (with limits). By increasing total bandwidth, the potential throughput is also increased, which is crucial for maximizing the throughput of one individual connection.
Another key benefit of MLO is the stability that comes with multiple channels. \cite{Deng20} With previous one-channel connections, if a client were to face any interference or loss, the device would either have to suffer the drop in performance or switch to a different channel, causing instability in both cases. \cite{Deng20} With MLO, even if one channel has a problem, the other channels can continue transmitting (or switch quickly when using non-simultaneous multi-link transmission), creating a much more seamless experience. \cite{Deng20}
Prior to MLO, many sections of the frequency band remained unused due to each device only being able to connect to one channel on one band. \cite{Deng20} Now that MLO can use all idle frequencies, these channels can be harnessed and not wasted. \cite{Deng20}

\subsection*{2.5.2 Limitations}
It’s important to note that WiFi access points have traditionally supported only a small number of links/outgoing channels for devices to connect; \cite{Ergen02} that is why there are so many methods in place to “share” the connection, because many devices have to contend for one outgoing AP link. \cite{BianchiMIMO} MLO is extremely effective for wireless networks that have only a few devices since the data-intensive user can take up a large share of airtime without significant consequences. \cite{Bellalta23} However, since MLO allows that high-demand user to utilize a larger share of the WiFi airtime, it forces the router to coordinate fairness mechanisms to properly share the link among all devices. \cite{Deng20}
Additionally, there are limits on the number of channels and number of bands based on hardware. \cite{Deng20} These limits are from both the access point and the device. MLO compatible APs are typically dual-band or tri-band, meaning two/three links (connections) can run at the same time. \cite{Deng20}

\section*{2.6 Many Devices}
Over time, the number of devices connecting to Wi-Fi has increased dramatically, \cite{Bellalta16} meaning routers have to handle more simultaneous requests from the new influx of devices \cite{Bellalta16}. Previous approaches, such as time and frequency division (Section 2.3) demonstrate how a WiFi connection can be shared among different users. Although these help routers divide their existing bandwidth among different users, it does not increase the throughput rates they have. The number of devices often causes the existing bandwidth to be insufficient. By the road analogy, the road space on the ground is limited; innovation can create more sophisticated ways to share the road, but we are uniquely bound to this width (bandwidth) of road. There is simply not enough space to accommodate the number of cars going to different customers on this road. So, how exactly can this problem be addressed?

\subsection*{2.6.1 MIMO}
The solution can be observed relatively intuitively when observing highways and roads. When two highways need to occupy the same width, a solution can be to simply build highways on top of each other. One highway sits at a lower level, while another highway is above in a double-decker fashion. Indeed, both highways cover the same 2-dimensional space (length and width-wise), but they occupy different spaces in the 3-dimensional plane (building straight up). This is essentially how the system works with Multi-Input Multi-Output or MIMO. MIMO uses spatial-division multiplexing (SDM), or transmitting parallel independent data streams, to multiply the total throughput potential. \cite{IEEE80211n}

\subsection*{2.6.2 SU-MIMO \& MU-MIMO}
Single-User MIMO or SU-MIMO was introduced in WiFi 4 (802.11n standard). \cite{IEEE80211n} This is MIMO/highway stacking concept except with only one user. The device has to go through normal contention to compete with other devices for the channel (2.3.2 CSMA),\cite{BianchiMIMO} but once the device has reserved the channel it can now split its traffic along each of the stacked highways. Each spatial stream can have data sent on it; under perfect/ideal conditions, the sending bitrate matches the number of streams (e.g., 8 streams and 8x bitrate). \cite{IEEE80211n} However, this was only available for one user at a time. \cite{IEEE80211n} Wi-Fi 5 (802.11ac) introduced Multi-User MIMO (MU-MIMO), fixing this problem. \cite{IEEE80211ac} This is incredibly effective for multiple simultaneous connections from different connections because the traffic can be routed at the same time while also increasing total throughput/bitrate. \cite{IEEE80211ac}

\subsection*{2.6.3 Limitations}
For MU-MIMO, just like SU-MIMO, devices must first contend for the connection; it is not 8 separate highways, but rather one highway that cars must merge onto, which then splits into 8 highways. Additionally, if clients are in the same physical space or in a similar direction relative to the AP, it can create interference which affects the signal of the spatial streams. \cite{IEEE80211ac}

Throughput benefits with real-world results vary widely with MIMO, with some cases actually causing a loss of 58\%. \cite{MuMimoMeas24} As a result, especially in crowded environments, some APs may have MU-MIMO disabled entirely. \cite{MuMimoMeas24}

In WiFi 6, MU-MIMO was developed to allow uplink as well. \cite{IEEE80211ax} WiFi 6 expanded on MU-MIMO with uplink spatial streams (in addition to downlink streams), allowing more streams from devices to APs, compared to WiFi 5’s streams on downlink (AP to device) only. \cite{MuMimoMeas24} With WiFi 8, the focus shifts away from expanding spatial streams and increasing raw data rate with “more features”, but rather refining existing features: one such example is WiFi 8’s goal to make the spatial streams more resistant to outside interference \cite{He25}

\section*{2.7 Peak Throughput}
In this WiFi development category, peak throughput through modulation will be addressed. This problem has ties to many of the previous sections, many of which will be addressed in the Limitations section (2.7.2). In order to get to the root of this problem, it’s important to understand how data is represented by radio waves.

\subsection*{2.7.1 Modulation}
Data is simply bits; 1s and 0s. A radio signal can be considered in its base form: a cycle (one period in the wave). For one cycle, a “1” can be considered a sine wave that begins at the origin (goes up, down, up) and a “0” can be considered a negative sine wave (down, up, down). With these two combinations, one bit can be represented. Now, one can take the same sine wave and shift its starting point pi/2 to the right. Do this for a total of 4 cycle representations, and those 4 representations can show all the possible combinations of two bits (00, 01, 10, 11).

Repeating this again, shifting the graph by pi/4 instead now nets 8 representations ($2^3$) for 3 bits. This can be continued for 16 representations ($2^4$) using 4 bits, and on. Alterations can be made to the amplitude of the wave as well, creating more opportunities for modulation and multiplying bitrate. \cite{Proakis01} This can be represented by a 2D map of amplitude and phase alterations and is known as constellation mapping. \cite{Proakis01} In recent years, this development of modulation has continued with each iteration of WiFi, with WiFi 6 specifying over 1,000 representations ($2^10$) \cite{Khorov19} over 10 bits and WiFi 7 recently supporting over 4,000 representations ($2^12$) over 12 bits. \cite{Deng20}.  This is analogous to packing more puzzle pieces in a box when a large jigsaw puzzle is ordered from Amazon. By shrinking the height of the puzzle pieces, we can fit more pieces in each box.  This process is known as Quadrature Amplitude Modulation or QAM, and having a single wave represent 12 bits of data is known as 4K QAM.

\subsection*{2.7.2 Limitations}
Although a potential 12x increase in transmission rates sounds groundbreaking, just like all other WiFi features it comes with limits. With the analogy, we can say the packages are significantly smaller in size due to the puzzle pieces being a fraction of their original height. While this does provide an opportunity to use smaller boxes and thus deliver significantly more puzzle pieces to the user, it also makes each puzzle piece much more fragile and much more easily damaged. With thicker puzzle pieces, slight disturbances or potentially even a minor car accident wouldn’t have shaken the car up enough to affect the puzzle pieces inside. However, thinner puzzle pieces do not have this luxury; even slight disturbances may destroy the puzzle piece beyond recognition. As such, if the road is prone to car accidents at a certain time, the delivery warehouse should avoid sending puzzle pieces that are sliced too thinly. Applying the analogy back to WiFi, it becomes clear that signal modulation has the potential to transmit data at a higher rate, but that rate comes at a cost; the signal must be consistently clean enough (interference-free) to allow the receiver to differentiate slight differences in signal. \cite{Proakis01}
Additionally, if interference is present in part of the band, even with wider channels (Section 2.1) and preamble puncturing (Section 2.4), the modulation drops across the entire channel. \cite{Halperin10} Signals look extremely similar at a higher Modulation and Coding Scheme (MCS), and they are extremely difficult to differentiate consistently if interference also exists; \cite{Bejarano16} even a small amount of interference can force a channel to decrease its MCS. \cite{Bejarano16} The modulation of a channel matches the modulation of the weakest link or the part most affected by interference, making wide channels extremely vulnerable to lower MCSs. \cite{Halperin10} Additionally, all MIMO spatial streams tied to a channel (Section 2.6) will have to drop to a lower MCS since they are all considered under the same channel. \cite{MCSSpatialStreams} Essentially, interference often affects the Modulation and Coding Scheme greatly and has a major impact on sending bitrate as a result. \cite{Bejarano16}
With WiFi 8, it is unlikely that there will be an expansion into 8K QAM, as the higher order modulation would require a much higher ratio of quality transmissions to noise, or signal-to-noise ratio (SNR). \cite{He25} If pursued, the total number of signal representations will need to be doubled, \cite{Proakis01} yet the transmission increase would be only slightly over 8\%.  The logarithmic relationship between number of signal representations and shows the diminishing returns with each iteration of QAM: (bits = log2(M) for M-QAM) \cite{Proakis01} hence, this is a likely factor to why the current in-progress WiFi 8 specifications are starting to shift away from a focus on pure rate-optimizing modulation and toward other methods, such as Unequal Modulation (UEQM). \cite{WiFi8Tutorial25} UEQM focuses more on modulation stability and retaining some modulation even in high-noise environments, which reflects WiFi 8’s focus toward UHR and consistency of connection. \cite{WiFi8Tutorial25,He25}

\end{document}